\documentclass{desyproc}
\usepackage{psfrag}
\usepackage{wrapfig}

\newcommand{\bb}{{\cal B}}
\newcommand{\am}{{\cal A}}

\renewcommand\({\left(}
\renewcommand\){\right)}
\renewcommand\[{\left[}

\newcommand{\be}{\begin{equation}}
\newcommand{\ee}{\end{equation}}
\def\bea{\begin{eqnarray}}
\def\eea{\end{eqnarray}}

\begin{document}
\title{Photon$\leftrightarrow$Axion conversions in transversely inhomogeneous magnetic fields}

\author{{\slshape Javier Redondo}\\[1ex]
Max-Planck-Institut, F\"ohringer Ring 6, D-80805 M\"unchen, Germany\\
}
\contribID{redondo\_javier}

\desyproc{DESY-PROC-2009-05}
\acronym{Patras 2009} 
\doi  

\maketitle

\begin{abstract}
We compute the photon-axion conversion probability in an external magnetic field 
with a strong transverse gradient in the eikonal approximation for plane waves. 
We find it typically smaller than a comparable uniform case.
Some insights into the phenomenon of photon-axion splitting are given.
\end{abstract}

Recently, a possible enhancement of photon$\leftrightarrow$axion conversions in magnetic fields with strong gradients has been proposed~\cite{Guendelman:2008ep,Guendelman:2009kv}.
Such a source of enhancement was searched in recent proposals to explain some aspects of the X-ray activity of the Sun: 
the longstanding corona problem and the triggering of solar X-ray activity~\cite{Zioutas:2008ie,Zioutas:2009bw}.
The core idea of these solutions is that axions created in the solar interior by the Primakoff effect, and therefore having energies corresponding to hard X-rays, could reconvert into photons in the outside layers of the Sun.

The standard mechanism for reconversion, inverse Primakoff, produces a too small deposition of energy to account for these effects. An alternative is the coherent conversion in the strong and long magnetic fields of the solar surface.
An estimate of this effects using a 1-D formula for the reconversion probability is not very promising either,  but
recent claims suggest that this might be not adequate and a full 3-D calculation could lead to surprises~\cite{Guendelman:2009kv,Zioutas:2009bw}. 
In particular, axion-photon conversion in strong magnetic field gradients has attracted some attention because could lead to an enhancement of the conversion probability due to the so-called ``photon-axion splitting''~\cite{Guendelman:2008jm}. 
If such an enhancement is realized, it could be very advantageous for laboratory experiments looking for axions, or axion-like-particles, to use quadrupole magnets instead of the usual dipoles. 
This possibility is already under consideration in the CAST and in the OSQAR ``light-shining-through-walls'' experiments at CERN~\cite{OSQAR_patras}.

In this contribution we present some simple calculations and physical insights on the phenomenon of photon-axion splitting. 
We find the 1-D estimation of the conversion probability to be reliable in the cases of interest. 
As a consequence, no enhancement is foreseeable in the solar environment and the use of quadrupoles presents no advantage over current dipoles.

\section{Invitation: Mirages}

A hot surface like a road in summer behaves as a mirror for objects and observers placed near to it. 
This phenomenon is called a ``mirage'' and has their origin on the curvature of light rays in the presence of a temperature gradient causing a gradient of the air's refraction index.

\begin{figure}[t]
\centering
{\psfragscanon
\psfrag{a}[][l]{$\nabla n \propto - \nabla T$}
\psfrag{b}[][l]{\hspace{7pt}$n$}
\includegraphics[width=13cm]{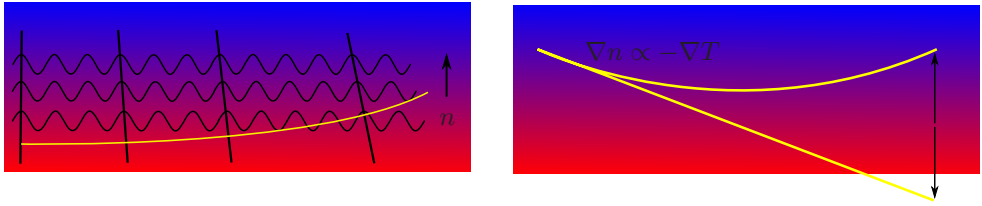}}
\vspace{-.3cm}
\caption{
\footnotesize
Qualitative evolution of a light wave propagating initially parallel to a hot surface, which forms a gradient of $n$.
The index of refraction increases with altitude, so the wave-number $k$ is increasingly larger with respect to the frequency $\omega$, making the phase-fronts increasingly closer with altitude. As the phase-fronts become tilted, the light rays, which are perpendicular to them, bend upwards forming a inverted image.}
\end{figure}

\section{The photon-axion system in a transverse gradient field}

In a transverse external magnetic field ($B$), the component of the axion field along the external magnetic field ($A$) mixes with the axion ($a$)  and the equations of motion can be written as~\cite{Raffelt:1987im}
\be
\left[
\square +
\( \begin{array}{cc} m_a^2 & g B \omega \\  g B \omega & 0 \end{array} \)
\right]
\( \begin{array}{c} A \\ a   \end{array} \) =0 \ .
\ee
Let us take $m_a=0$ for the purpose of illustration. 
In this case, the equations can be diagonalized by considering the linear superpositions $A_\pm=(A\pm a)/\sqrt{2}$ which therefore evolve according to
\be
\square A_\pm = m_\pm^2 A_\pm \quad\quad;\quad\quad {\rm with}\quad\quad m_\pm^2 = \pm g B \omega \ .
\ee
The $A_\pm$ waves have indices of refraction with opposite sign given by $n_\pm^2=1-m^2_\pm/\omega^2=1\mp g B/\omega$ and
therefore, in complete analogy with the mirage effect, the $A_\pm$ rays will curve in {\em opposite} directions if there is a transverse gradient of the magnetic field. 
This leads to the so-called ``photon-axion splitting''~\cite{Guendelman:2008jm} with very speculative and spectacular consequences~\cite{Chelouche:2008ax}.

\begin{figure}[t]
\label{fig2}
\hspace{.5cm}
{\psfragscanon
\psfrag{c1}[][l]{$B=\rm const.$}
\psfrag{c2}[][l]{$\nabla B= \rm const.$}
\psfrag{a}[][l]{$\lambda$}
\psfrag{b}[][l]{$g B \lambda /\omega$}
\psfrag{x}[][l]{$x$}
\psfrag{z}[][l]{$z$}
\includegraphics[width=5cm]{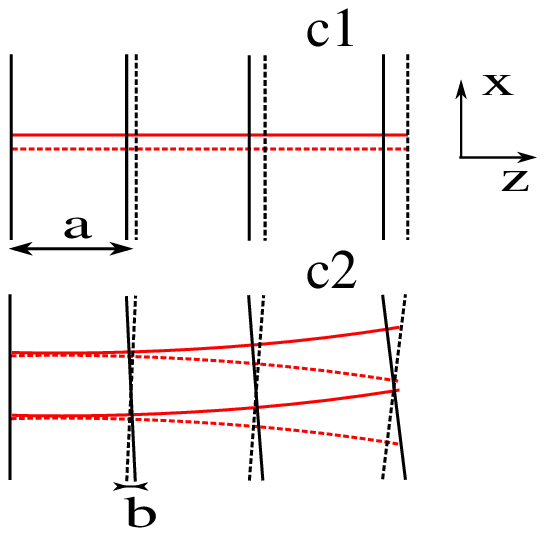}}
{\psfragscanon
\psfrag{x}[][l]{$x$}
\psfrag{y}[][l]{$y$}
\psfrag{z}[][l]{$z$}
\psfrag{B0}[][l]{$\vec B_{\rm ext}=0$}
\psfrag{B}[][l]{$\vec B_{\rm ext}(x)$}
\psfrag{Bg}[][l]{$\vec \nabla B_{y,\rm ext}$}
\psfrag{Bgg}[.8][l]{\hspace{1cm}$\color{red} \nabla n_\pm \propto \mp  \nabla B_{y,\rm ext}$}
\psfrag{A+}[][l]{$A_+$}
\psfrag{A-}[][l]{$A_-$}
\includegraphics[width=8cm]{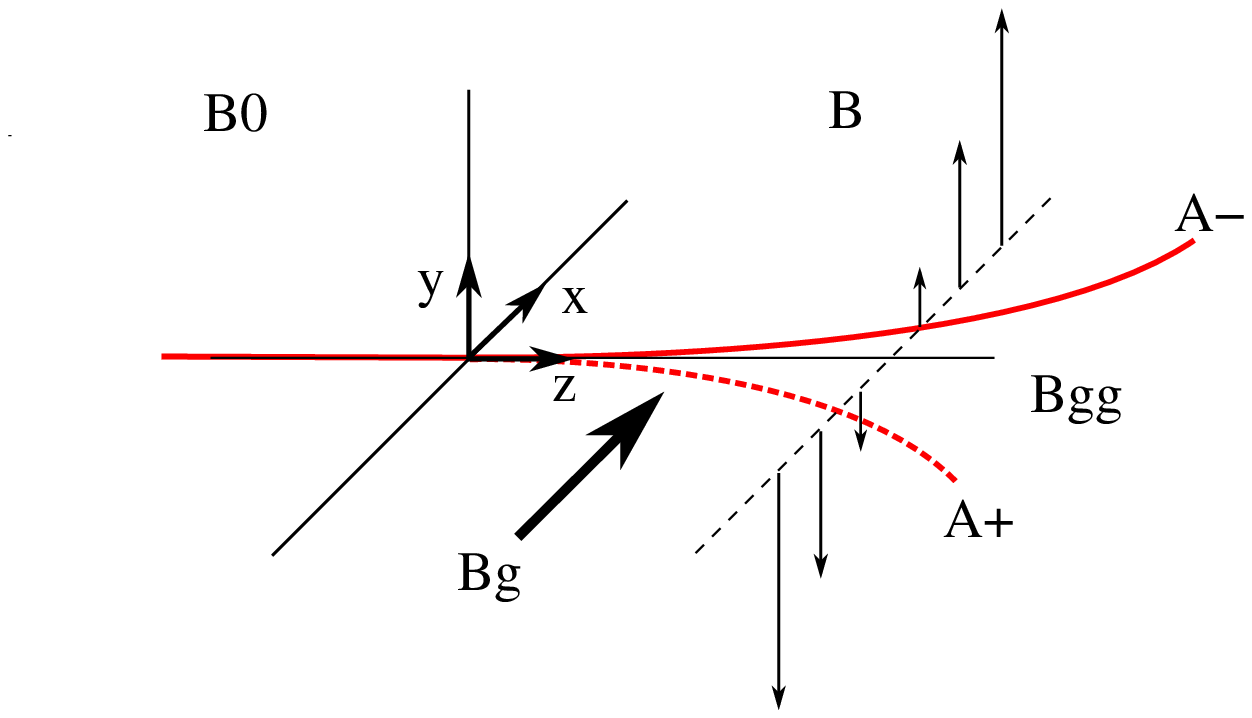}}
\caption{
\footnotesize 
(LHS) Evolution of the $A_\pm=(A\pm a)/\sqrt{2}$ phase fronts (black dashed and solid,respectively) in the presence of a  transversely homogeneous (up) and constant gradient (down) magnetic field. The corresponding rays are shown in red.
(RHS) The gradient case shown in perspective.
}
\end{figure}

We can get a non very sophisticated first look at the photon$\leftrightarrow$axion conversion probability by looking at the evolution of $A_\pm$ phase fronts. In Fig.~\ref{fig2} we can see a comparison of the homogeneous and constant gradient case.
For this first look we neglect any diffraction effect and changes on the $A_\pm$ amplitudes, we simply consider a 1-D problem for each value of $x$.
The phase fronts of $A_\pm$ separate a distance $(n_+^{-1}-n_-^{-1})\lambda_0\simeq g B \lambda/\omega$ in a wavelength $\lambda_0$. 
Equivalently, after a distance $Z$ there is a phase difference of $\phi_+-\phi_-\simeq - g B Z$ between the $A_+$ and $A_-$ waves.
This phase difference is constant in the $x$ direction in the homogeneous case but increases linearly for a constant gradient.

The $\gamma\leftrightarrow a$ conversion probability per unit transverse length can be compared in both cases. 
In each case it can be written as $c\| e^{\i \phi +}- e^{\i \phi -}\|^2/4 = c\sin^2 (g B Z/2)$ 
with $B=B_0$ for a constant field and $B=B_1 x$ for the gradient case ($c$ is a normalization unimportant for our purposes).  
The ratio of the two probabilities integrated over an interval $x\in (-X,X)$ is then
\bea
\frac{\left. P\right|_{ B=B_1x}}{\left. P\right|_{ B=B_0}} = \frac{\frac{1}{2}\(1-{\rm sinc}(2g B_1 X Z)\)}{\sin^2(g B_0 Z/2)}\xrightarrow{g B Z \ll 1} \frac{1}{3}\(\frac{B_1 X}{B_0}\)^2
\eea

The last limit is relevant in usual practical applications, given the smallness of the values of the axion-photon coupling $g$ allowed by stellar evolution arguments as well as the typical sizes and strengths of magnetic fields.
Note that $B_1X$ gives the maximum magnetic field in the $x$-interval we have used. 
Superconducting quadrupoles have their field gradients limited precisely by the critical field at the boundaries, so in comparing a quadrupole with a similar dipole one should use $B_1 X \sim B_0$. 
Quadrupoles are therefore less efficient than dipoles. 
This result is very easy to understand.
By making the problem 1-D, only the strength of the magnetic field squared matters, which in average is of course smaller in a quadrupole than in a dipole. 

\section{Solution in the eikonal approximation}

The most important 2-D features of the problem can be studied in the eikonal approximation. 
Starting with the ansatz $A_\pm(x,z,t)={\cal A}_\pm(x,z) e^{-\i \omega t}e^{\i \omega S_\pm(x,z)}$ the equations of motion become
\be
\label{ansatz2}
(\nabla S_\pm)^2-\i \frac{1}{\omega} \frac{\nabla \(\am_\pm^2 \nabla S_\pm\)}{\am_\pm^2}-\frac{1}{\omega^2}\frac{\nabla^2\am_\pm}{\am_\pm^2} = 1-\frac{m^2_\pm}{\omega^2}\equiv n_\pm^2(x)
\ee
The eikonal approximation amounts to neglect the second and third terms of the LHS, which are suppressed if diffraction effects occur in length scales much longer than the wavelength $\lambda_0$. We will solve only the equation of $A_+$ since the solution of $A_-$ is given by the former by changing the sign of the magnetic field. 
The eikonal equation $(\nabla S)^2=n^2(x)$ can be solved by the method of characteristics, i.e. finding a 
one-parameter family of curves $\vec r_{x_0}(s)=(x(s),z(s))$ (we drop the trivial component $y$) that satisfy the Hamilton equations
\be
\label{ray}
\frac{d\vec r}{ds} = \vec p
\quad ; \quad
\frac{d\vec p}{ds} = \vec \nabla n^2/ 2 \ . 
\ee
with initial conditions at the boundary of the magnetic field $\vec r(0)  = (x_0 ,  0), \vec p (0) = ( 0 , n_0 )$ with $n_0=n(x_0)$.
These are given by
\be
\label{ray:sol}
\vec r  (s,x_0) = \left( x_0 -\bb s^2/4 ,  s n_0 \right)
\ee
where we have defined $\bb = g B_1/\omega$, the quantity that controls the inverse of the radius of curvature of rays 
$R = 4 \bb^{-1} (1+(\bb s/2)^2)^{3/2}$, which turns out to be \emph{huge} for typical parameters 
\be
\bb^{-1} = \frac{\omega}{ g B_1} \simeq 5 \times 10^{17}{\rm m} \(\frac{\omega}{\rm keV}\) \(\frac{g}{10^{-10} {\rm GeV}^{-1}}\)^{-1}
\(\frac{B_1}{{\rm T/cm}}\)^{-1}
\ee
The angle of divergence of the $A_\pm$ rays after a length $z$ is $\theta_\bb\simeq \bb z$.
 
The eikonal function giving the evolution of the $A_\pm$ phases is given by 
\be
S=\int_0^s \vec\nabla S \cdot d\vec r(s) = s n^2_0 + \frac{\bb^2 s^3}{12} = z n(x) \(\frac{2}{1+\sqrt{1-\xi^2}} \)^{1/2}  \(\frac{2+\sqrt{1-\xi^2}}{3}\) \ , 
\ee
where $\xi=\bb z /(1-\bb x)$ is used. In Fig. 3 we show some rays and iso-contours of the eikonal function $S$.
Defining $\Delta S(x,z,\bb)= S(x,z,\bb)-S(x,z,-\bb)$, the photon-axion conversion probability per unit transverse length in the eikonal approximation is
\be
\frac{dP_{_{2D}}/dx}{dP_{_{1D}}/dx} = \frac{\sin^2 \omega \Delta S/2}{\sin^2 gBz /2} \xrightarrow{\bb x<\bb z\ll 1} 
\frac{\sin^2 \(g B_1 x z/2 \(1+(\bb z)^2/8\)\)}{\sin^2 gBz /2} . 
\ee
Where we have normalized again to the 1-D result for comparison.
Note that this is a very small correction to our previous 1-D rough estimate if $\bb z \ll 1$. 

So far we have considered only infinite plane waves. 
If our photon or axion beam passes through a confined region of size $X$, it will suffer diffraction with a characteristic angle given by
$\theta= 1.22 \lambda_0/X$ which might be larger than the splitting angle if $\bb z (\omega X) =g B_1 X z \ll 1$ (the typical case except maybe in very extreme conditions~\cite{Chelouche:2008ax}).
However, this certainly does not evade our conclusion when diffraction is negligible, which would occur for an helioscope using a quadrupole magnet. 
Indeed it is very likely that even beams with large diffraction don't show additional enhancements in the photon$\leftrightarrow$axion probability either. 
We have calculated the conversion probabilities for typical laser beam parameters used in light-shining-through-walls experiments and found no surprise at all. These results will be presented elsewhere~\cite{JaecRed}

\begin{wrapfigure}{r}{60mm}
{\psfragscanon
\psfrag{a}[][l]{$\bb z$}
\psfrag{b}[][l]{$\bb x$}
\includegraphics[width=5.5cm]{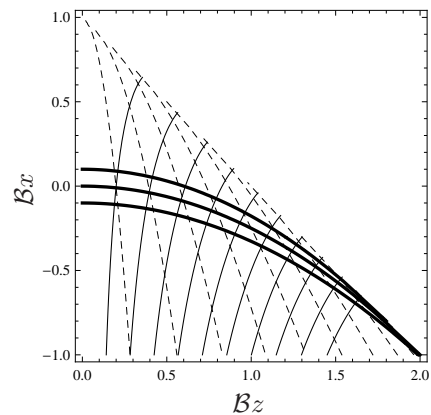}}
\caption{
\footnotesize 
Rays (thick), phase fronts (thin) and iso-contours of optical path $s$ (dashed) of the field $A_+=(A+a)/\sqrt{2}$. 
A caustic is formed by the accumulation of rays in $\xi=1$, ($\bb z=1-\bb x$),  the wave cannot propagate in the upper region. 
The length scales are normalized to the characteristic length scale of the ray deflection $\bb^{-1}$.
The evolution of $A_-$ is the mirror image with respect to the $x=0$ line.
}
\end{wrapfigure}

\section*{Acknowledgements}

I would like to thank the organizers of this very interesting meeting, and specially to the local committee, for having helped notably in creating a beautiful atmosphere and for taking care marvelously of all of us throughout the workshop.


\begin{footnotesize}

\providecommand{\href}[2]{#2}\begingroup\raggedright\endgroup




%

\end{footnotesize}


\end{document}